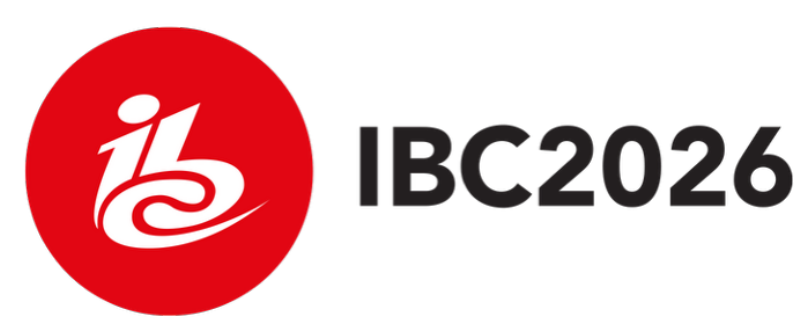

# Scalable SSIM Estimation from PSNR for Per-Title and Context-Adaptive Encoding Workflows

L. Trudeau[1] and M.G. Martini[2]

[1]Université du Québec à Rimouski, Canada
[2]Kingston University London, UK

## ABSTRACT

Modern streaming pipelines run hundreds of candidate encodes per asset to support per-title encoding, shot-based optimization, and context-adaptive ABR ladder construction. These techniques have moved perceptual quality metrics into the critical path: SSIM and VMAF now guide encoding decisions rather than passively monitor them. We measure that SSIM evaluation accounts for 7–35% of x264 encode time at production speed presets, with the cost ratio rising as encoders run faster. We propose ApproxSSIMate, a low-complexity method for estimating SSIM from PSNR combined with reference-sequence statistics computed once per sequence and reused across every candidate encode. This decouples quality estimation from the encode-decode-compare loop, enabling perceptual quality feedback in live encoding and amortizing quality measurement across candidate encodes in per-title workflows. We validate the approach across H.264/AVC, H.265/HEVC, and AV1 on the Objective-1-fast dataset and release the implementation as free and open-source software.

## Introduction

Streaming a single feature film today involves upwards of a thousand encodes. Whether for live events or video-on-demand, pipelines segment videos into chunks ranging from a fraction of a second to ten seconds. Each chunk is encoded into tens of candidate encodes that vary in bitrate and resolution.

Pipelines score each encode with perceptual quality metrics such as SSIM, introduced by Wang et al (1), or VMAF, proposed by Li et al (2). Content-aware encoding (CAE), described by Aaron et al (3), elevated these metrics from passive monitoring to active control. Perceptual quality metrics aren't free. As shown later in Cost of SSIM in Encoding Pipelines, SSIM evaluation accounts for 7–35% of encode time on x264, with the ratio rising at faster encoder presets. Encoders have decades of optimization work behind them and offer a wide range of speed presets; quality metrics, despite now sitting in the critical path of video pipelines, have received comparatively little attention as an optimization target.

Figure 1 illustrates the architecture of a modern encoding pipeline. For each candidate encode, the encoded frames must be decoded, and the reconstructed frames must be compared against the uncompressed source frames using a full-reference quality metric such as SSIM. The resulting quality measurements are then fed back to the content-adaptive optimizer, which selects or adjusts the encoding parameters for subsequent encodes. Although this loop enables more efficient bitrate ladders, it also places quality evaluation in the critical path of the encoding workflow. As the number of chunks, renditions, and candidate encoder configurations increases, the cost of repeatedly decoding, transporting, and evaluating reconstructed frames becomes a meaningful part of the overall computation.

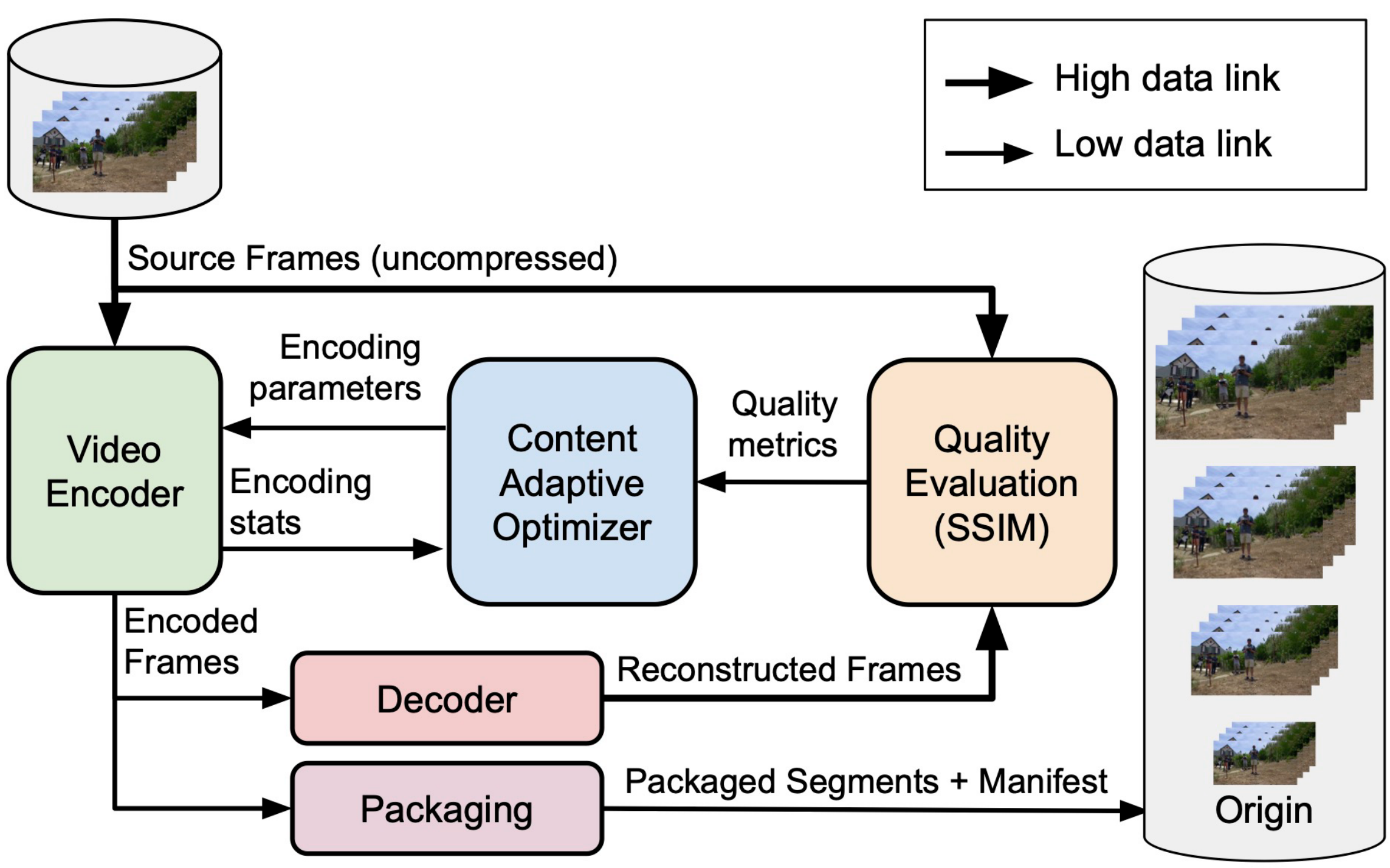

Figure 1 – Conventional content-adaptive encoding workflow using full-reference quality evaluation. Each candidate encode must be decoded and compared against the uncompressed source frames using SSIM before quality metrics can be fed back to the content-adaptive optimizer. Thick arrows indicate high-data links carrying video frames, while thin arrows indicate low-data links carrying parameters, statistics, or quality scores.

Martini (4) observed that SSIM decomposes into a source-dependent and a distortion-dependent part and used this decomposition to derive a fast approximation for still images. In this paper, we show that the underlying assumptions also hold for compressed video sequences and evaluate a modified sequence-level model on video datasets. We show that the source-dependent term can be reused across multiple encodes of the same content, and further that it reduces to a single reference-derived constant per sequence. This enables sequence-level SSIM approximation from average PSNR alone.

We call this method ApproxSSIMate. It requires only the PSNR of the encoded sequence and the reference-derived constant, which can be computed in parallel with the encoding. This is particularly valuable for live-streaming pipelines where post-hoc quality measurement is not an option. The implementation is released as open-source software by Trudeau and Martini (5), and we validate it across H.264/AVC, H.265/HEVC, and AV1.

The remainder of this paper reviews related work, quantifies the cost of SSIM evaluation, validates the local-statistics assumptions, presents the ApproxSSIMate workflow, evaluates it across modern video encoders, and discusses practical integration into per-title and context-adaptive encoding workflows.

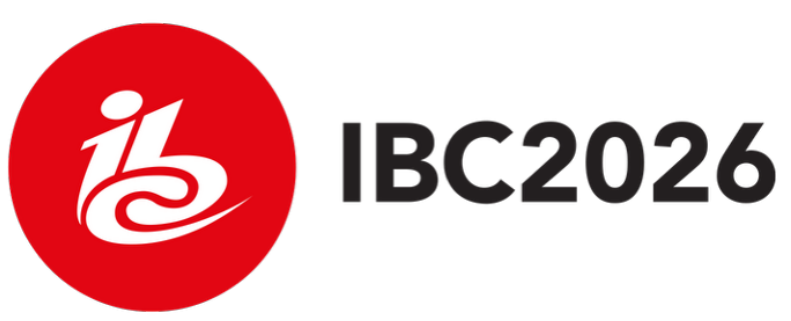

## RELATED WORK

Early attempts to estimate SSIM based on PSNR in multimedia systems were based on conversion tables (Zinner at al, 6). These tables ignored image and video content and relied on empirical threshold values. In the context of visual distortion resulting from packet loss, Reibman et al. (7) proposed a simple relationship for block-based SSIM as a function of MSE, information from the original and impaired images, and constant terms.

The works from Horé and Ziou (8, 9) further analysed the relationship between PSNR and SSIM. The authors provide analytical expressions and approximations for different use cases, requiring the joint processing of the original and compressed image, beyond what is required for PSNR. As part of their discussion on the validity of SSIM as a quality metric, Dosselmann et al. (10) showed that the index is directly related to the mean squared error. They compared the two metrics statistically and derived a pair of functions that algebraically connect them via the means and mean-square values of the original and impaired images. Their formulation requires information from both the original and impaired images in addition to MSE.

Wang et al. (11) examined the relationship between MSE and SSIM as optimization objectives, highlighting that the two are related but not interchangeable. This supports the motivation for methods that retain the efficiency of MSE-based measurements while better approximating SSIM behaviour.

After recognizing that the SSIM expression can be simplified to a correlation coefficient when $\mu_x = \mu_y$, Palubinskas (12) proposed an alternative quality metric based on similar statistics. Other authors, such as Shanableh (13), aimed to estimate SSIM without access to the reference, using bitstream and/or reconstructed-video information. These approaches compare SSIM estimation with PSNR estimation, but do not establish a direct relationship between the two metrics.

Bounds on the SSIM index for compressed images are provided by Channappayya et al (14) as a function of quantization rate for uniform, Gaussian, and Laplacian sources. The authors propose using these bounds for rate allocation problems in practical image and video coding applications. No relationship with PSNR or MSE is established.

Tan et al (15) developed a perceptually relevant MSE-based image quality metric. In doing so, they assume an additive error model and independence between signal and error. This metric was also adopted by Yeo et al (16) and used as a benchmark in the results section of Martini (17). It was shown that this estimation is less accurate than the methods proposed for Discrete Cosine Transform (DCT)-compressed images. Similar limitations in estimation accuracy were also reported in Li et al (18).

The mathematical properties of the structural similarity index have been studied in Brunet et al (19). More recently, Reznik (20) proposed an alternative model of SSIM computation using subband decomposition and identical distance measures in each subband, and discussed the relationship with PSNR, highlighting that the two are linked through a joint statistic of both signals, namely covariance.

## COST OF SSIM IN ENCODING PIPELINES

How expensive is SSIM in practice? This section quantifies the cost of SSIM evaluation in production-grade encoding pipelines. For each test sequence and encoder configuration, we measure encoding time and SSIM computation time independently.

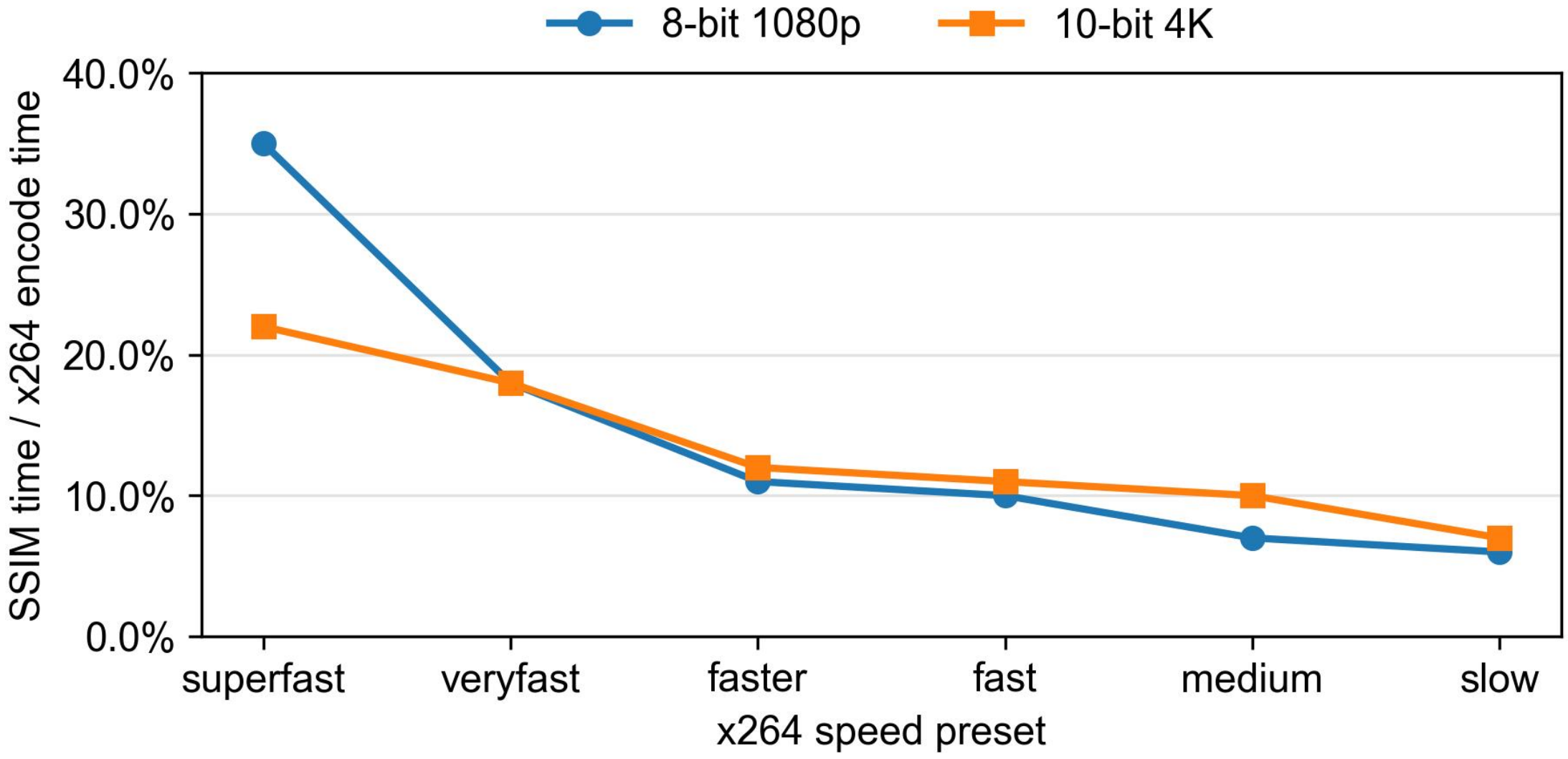

Figure 2 – SSIM cost rises sharply with encoder speed. At streaming-grade presets, SSIM accounts for 7–35% of x264 encode time, with the cost ratio growing as encoders run faster. Median of three FFmpeg runs over 12 1080p 8-bit sequences and 7 4K 10-bit Netflix sequences.

All measurements were performed using FFmpeg, the de facto tool for streaming encoding workflows. We used the x264 encoder for compression and the SSIM filter for quality measurement. The experiments were run on an AMD Ryzen 9 8945HS, which supports both AVX2 and AVX-512, the full range of SIMD optimizations available to FFmpeg.

We use 12 1080p 8-bit sequences and 7 4K 10-bit sequences from the Netflix Open Content library (21), chosen to span a range of spatial and motion complexity. To reduce noise in the time measurements, we ran each measurement three times and reported the median. Figure 2 shows SSIM computation time as a percentage of x264 encode time for the speed presets: superfast, veryfast, faster, fast, medium, and slow. These presets span the operational range from low-latency streaming through video-on-demand.

The trend is clear: SSIM cost as a fraction of encode time rises monotonically with encoder speed. At fast presets, the regime in which most cloud streaming pipelines operate, SSIM accounts for 11–18% of compute. At superfast, used for low-latency live encoding, the ratio reaches 22–35%. Across the full range, SSIM evaluation is a non-trivial fraction of pipeline cost, and the cost grows as pipelines optimize for throughput.

### EMPIRICAL VALIDATION OF ASSUMPTIONS

Martini (17) derived the local-MSE SSIM approximation under two assumptions about how compression affects local pixel statistics: that local-window means (A1) and local-window second moments (A2) are preserved between source and compressed images. We show empirically that these assumptions hold approximately for three encoders representative of modern production pipelines: x264, x265, and SVT-AV1.

Each sequence in the Objective-1-fast dataset was encoded across a QP sweep covering the operational range of streaming applications. Figure 3 summarizes the empirical validation of the two local-statistics preservation assumptions for x264, x265, and SVT-AV1. The top row measures A1 using the relative local-mean difference, while the bottom row measures A2 using the relative local second-moment difference. Since variance can be written as $\sigma^2 = E[x^2] - \mu_x^2$, measuring second-moment preservation under A1 is equivalent to validating variance preservation.

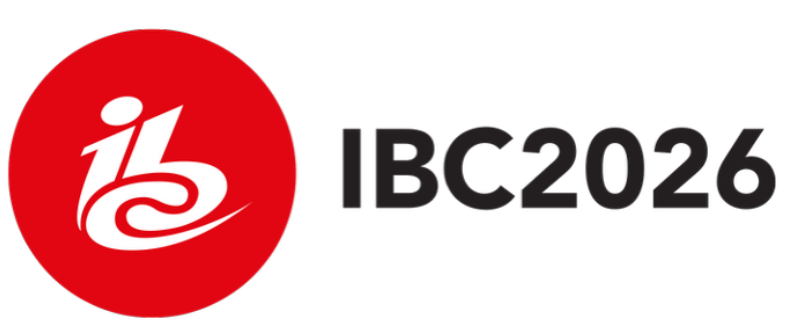

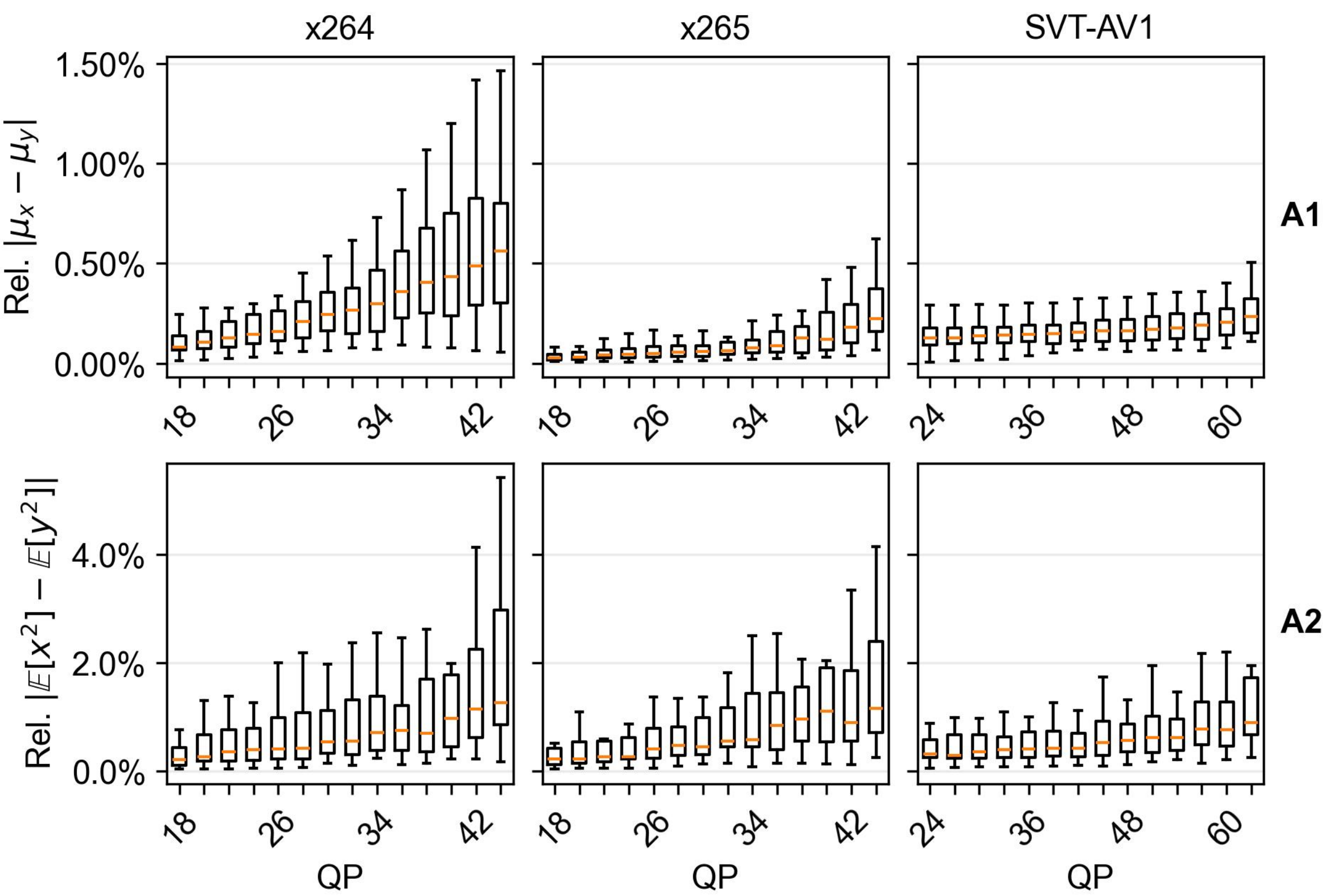

Figure 3 – Empirical validation of the local-statistics preservation assumptions on the Objective-1-fast dataset for x264, x265, and SVT-AV1. The top row reports the relative local-mean difference between source and reconstructed windows, testing assumption A1. The bottom row reports the relative local second-moment difference, testing assumption A2. Across encoders, local means remain tightly preserved, while second moments show greater spread as compression becomes more aggressive.

The larger deviations observed for x264 are consistent with its lower reconstruction quality relative to x265 at comparable QP values. Since the local-statistics preservation assumptions degrade as compression becomes more aggressive, this helps explain why both the mean and second-moment differences are larger for x264 in Figure 3.

Across all three codecs, A1 holds tightly: $\frac{|\mu_x - \mu_y|}{\mu_x}$ stays below 1% at the 95th percentile across the full QP range for x265 and SVT-AV1, and stays below 2% for x264. A2 shows greater spread, with the median relative second-moment difference staying below 2% within the operational QP range and per-sequence outliers reaching upward of 8% at extreme QPs. Consistent with Martini's still-image analysis, both assumptions hold most tightly at high quality and degrade as compression becomes more aggressive.

### PROPOSED APPROACH

Instead of computing full local-window SSIM between the source and reconstructed frames for every candidate encode, the proposed workflow computes reference-side statistics once for the source sequence. For each encoded rendition, the only required distortion information is the sequence-level mean squared error (MSE), or equivalently PSNR. These quantities are already available in many commercial encoders and quality-analysis pipelines.

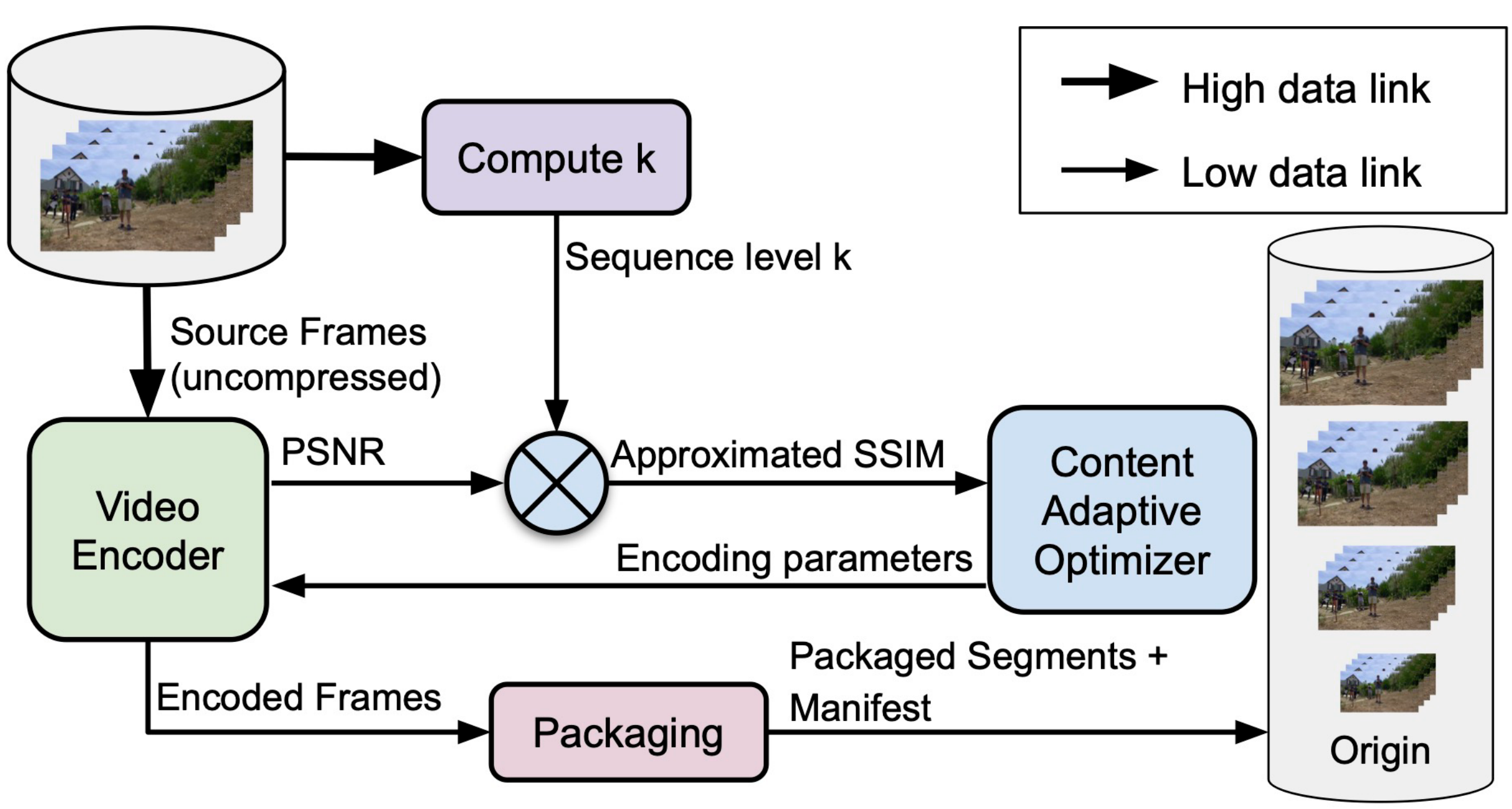

Figure 4 – Proposed ApproxSSIMate workflow for content-adaptive encoding. A reference-derived sequence constant is computed once from the uncompressed source frames and reused across candidate encodes. For each encode, the encoder-reported PSNR is combined with to estimate sequence-level SSIM, allowing quality feedback to be provided to the content-adaptive optimizer without repeatedly decoding and evaluating reconstructed frames using full local-window SSIM. Thick arrows indicate high-data links carrying video frames, while thin arrows indicate low-data links carrying parameters, statistics, or quality scores.

The approximation builds on the local relationship between SSIM and MSE established by Martini (17). Under the assumptions validated in Section IV, local SSIM can be approximated from local distortion and reference-side local variance. In this work, we consider the more restrictive practical setting where local distortion is unavailable and only the global sequence-level distortion is known. We therefore redistribute the global MSE across local windows according to reference-side spatial activity.

Empirically, direct variance-based redistribution over-allocates distortion to highly textured regions. We therefore use local standard deviation to obtain a softer activity weighting. This leads to a sequence-level approximation of the form

$$\widehat{MSSIM}(X,Y) = 1 - k_X \cdot MSE_G(X,Y) \tag{1}$$

where $MSE_G(X,Y)$ is the mean squared error over the sequence and $k_X$ is a source-dependent constant computed from the reference sequence.

For a sequence with $M$ local windows, $k_X$ is computed as

$$k_X = \frac{1}{M}\sum_{j=1}^{M} \frac{\sigma_{x,j} + \epsilon}{E[\sigma_x + \epsilon]\left(2\sigma_{x,j}^2 + C_2\right)}, \tag{2}$$

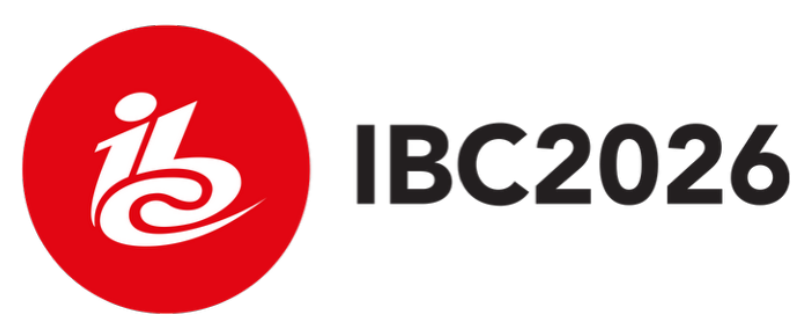

where $\sigma_{x,j}$ is the local standard deviation of the reference signal in window $j$, $C_2$ is the standard SSIM stabilization constant, and $\epsilon$ is a small numerical constant. The normalization term $E[\sigma_x + \epsilon]$ ensures that the redistributed local distortion averages back to the global MSE. When PSNR is used instead of MSE, the sequence-level MSE is obtained as

$$MSE_G = L^2 \cdot 10^{-\frac{PSNR}{10}}, \tag{3}$$

where $L$ is the number of levels used to represent a pixel value, $L = 2^B$, where $B$ is the number of bits per pixel per component (e.g., luminance).

Figure 4 summarizes the resulting workflow. The source sequence is analysed once to compute $k_X$. Each candidate encode then provides PSNR or MSE, which is combined with $k_X$ to estimate sequence-level SSIM. This removes the need to repeatedly compute full local-window SSIM between the source and reconstructed frames for every candidate encode.

## RESULTS

To evaluate ApproxSSIMate on video content, we used the Objective-1-fast dataset (22). Each sequence contains 60 frames. Quality metrics were computed on each frame and then averaged at the sequence level. We selected x264 (23), x265 (24), and SVT-AV1 (25) as representative open-source encoders for H.264/AVC, H.265/HEVC, and AV1, respectively. Each sequence was encoded over a 14-point QP sweep chosen to cover a broad range of operating points, from high-quality encodes to aggressive low-bitrate encodes. For x264 and x265, QP values ranged from 18 to 44 with a step of 2. For SVT-AV1, QP values ranged from 24 to 63 with a step of 3, reflecting the different quantizer scale used by SVT-AV1.

SSIM values were computed using the implementation provided by the scikit-image library (26). With the default settings used in these experiments, scikit-image computes SSIM using local 7 × 7 window-based statistics following the formulation introduced by Wang et al. (1). This implementation was chosen because it is widely used, reproducible, and directly available in scientific Python workflows.

Figure 5 reports the aggregate sequence-level SSIM results averaged across the Objective-1-fast dataset for x264, x265, and SVT-AV1. ApproxSSIMate follows the measured mean SSIM closely over most of the tested QP range. The same error pattern is observed for all three encoders: a small positive bias at high quality, followed by a crossover in the high-QP range, after which ApproxSSIMate increasingly underestimates the measured mean SSIM.

The largest deviations occur for the most aggressive compression settings. However, the aggregate curves remain monotonic and preserve the overall degradation trend as QP increases. This suggests that ApproxSSIMate is well suited for low-cost trend estimation and operating-point comparison, especially outside the very-low-quality regime.

Figure 6 shows the distribution of the per-sequence approximation error at each QP value for x264, x265, and SVT-AV1. The error is defined as ApproxSSIMate minus measured mean SSIM. These plots complement the aggregate curves by showing how the approximation behaves across individual sequences rather than only in the dataset average.

For all three encoders, the error distribution remains narrow and centred close to zero at low and moderate QP values. The median error shows a small positive bias in the middle QP range before shifting below zero at lower quality. As QP increases, the interquartile range becomes larger and the median shifts below zero, indicating that the approximation error

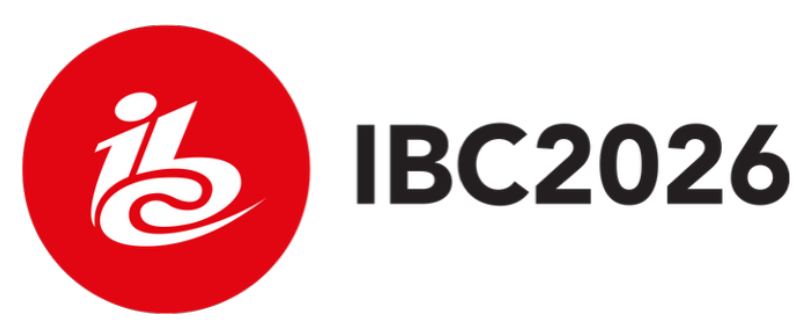

becomes more content-dependent and increasingly negative under stronger compression.

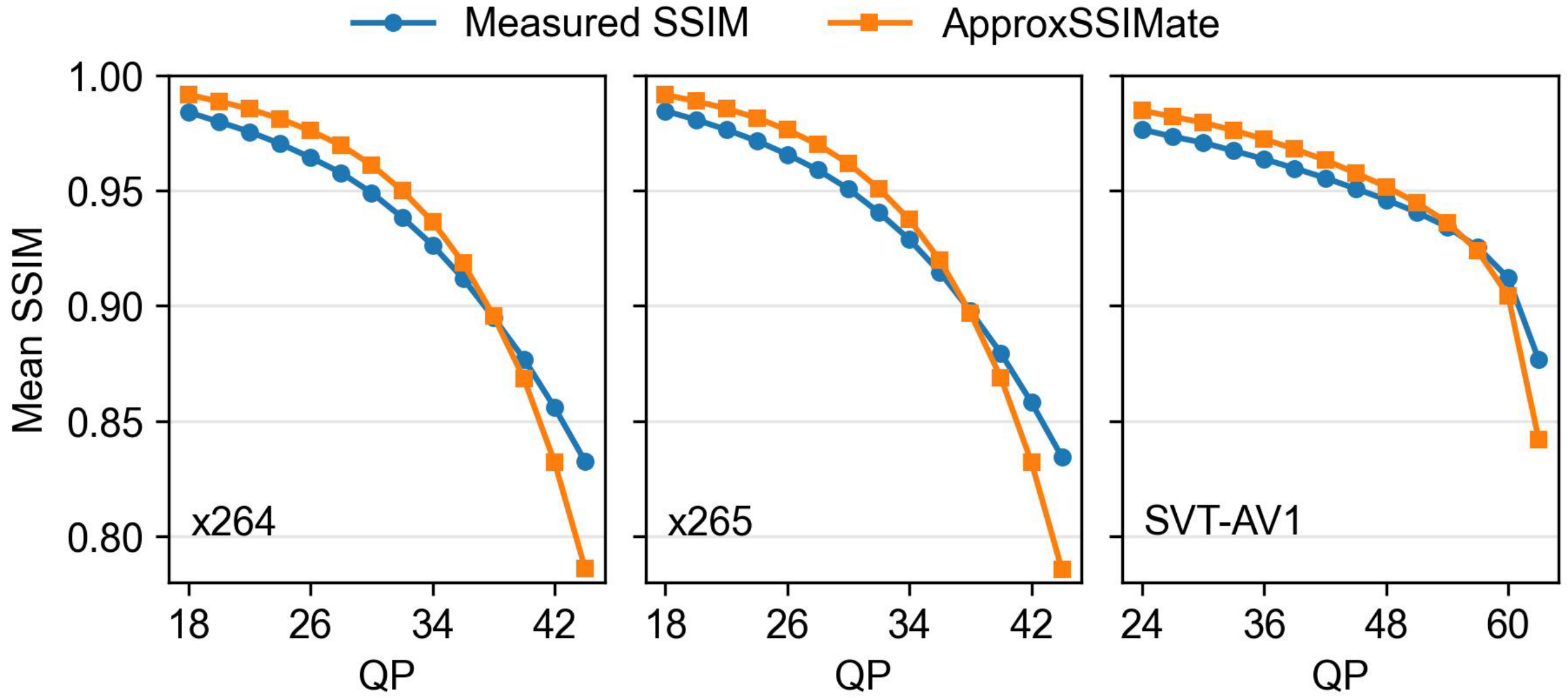

Figure 5 – Aggregate sequence-level SSIM comparison on the Objective-1-fast dataset for x264, x265, and SVT-AV1. ApproxSSIMate tracks measured mean SSIM closely across most of the tested quality range for all three encoders. At low QP values, the approximation slightly overestimates measured SSIM; at high QP values, it underestimates measured SSIM. Results are averaged across 30 sequences.

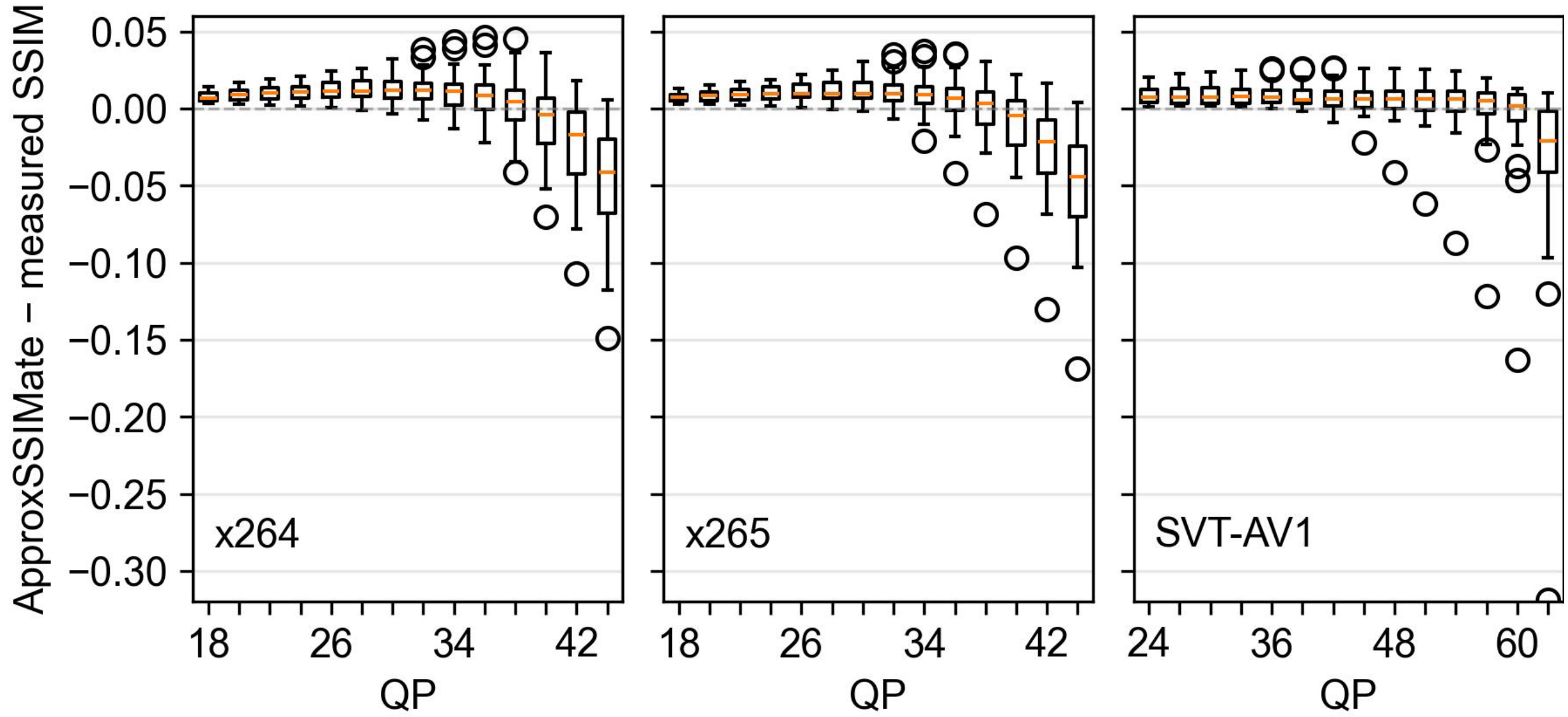

Figure 6 – Distribution of per-sequence ApproxSSIMate error on the Objective-1-fast dataset for x264, x265, and SVT-AV1. Error is defined as ApproxSSIMate minus measured mean SSIM. Across all three encoders, the distribution remains narrow and centred close to zero at low and moderate QP values, then widens and shifts toward negative error at high QP values.

The same qualitative behaviour is observed for x264, x265, and SVT-AV1. This suggests that the main source of approximation error is not specific to a particular encoder, but is instead associated with the interaction between content characteristics and the distortion regime. In particular, the largest deviations occur at the highest QP values, which is consistent with the divergence observed in the aggregate curves.

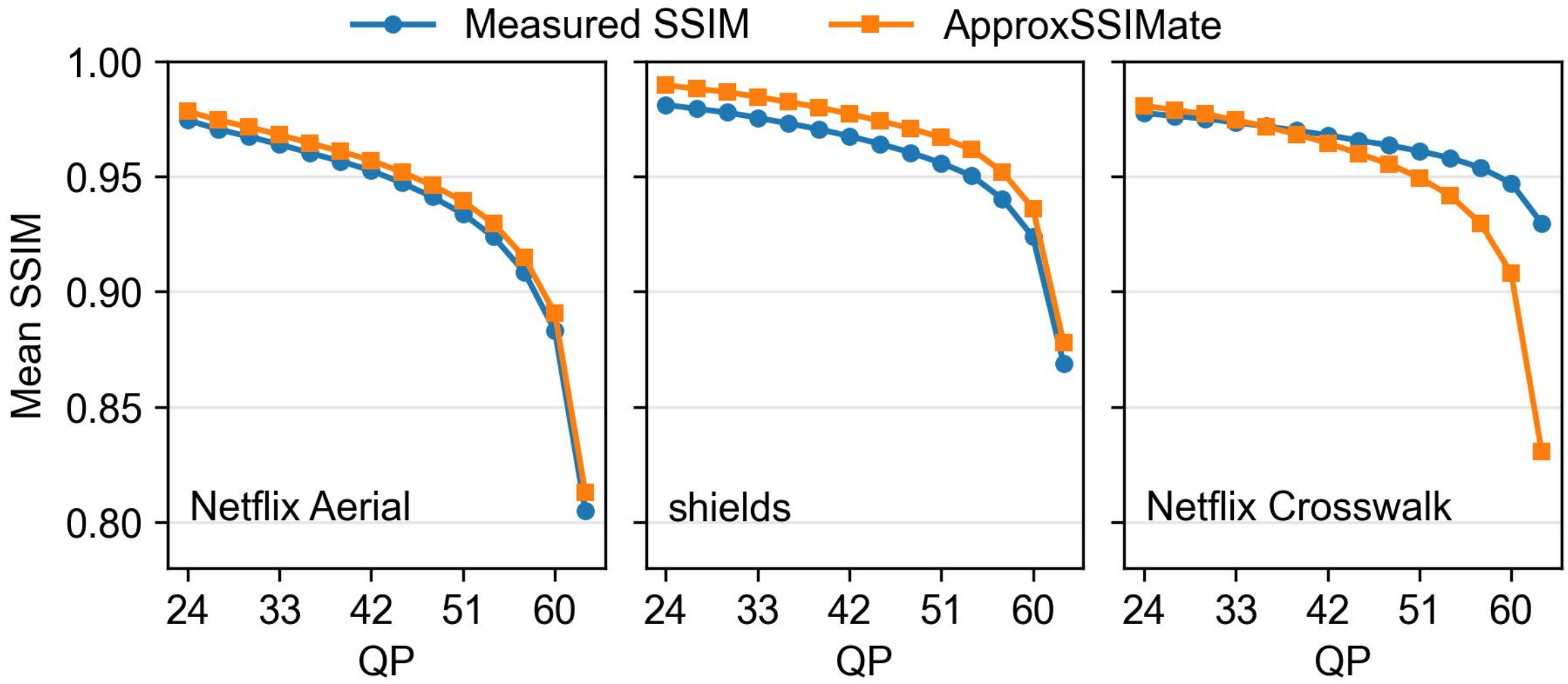

Figure 7 – Representative sequence-level SSIM comparisons for SVT-AV1 on the Objective-1-fast dataset. The selected sequences span the observed approximation behaviour: *Netflix Aerial* shows very close agreement between measured SSIM and ApproxSSIMate, *shields* represents an intermediate case, and *Netflix Crosswalk* is among the most challenging sequences. Even in the challenging case, ApproxSSIMate preserves the overall degradation trend, although it increasingly underestimates measured SSIM at high QP values.

The boxplot analysis describes the distribution of approximation errors, but does not show how these errors appear at the sequence level. Since the aggregate and error-distribution results show similar behaviour across encoders, Figure 7 uses SVT-AV1 as a representative modern encoder to illustrate individual sequence behaviour in more detail.

For Netflix Aerial, ApproxSSIMate closely follows both the slope and absolute value of the measured SSIM curve. The shields sequence shows an intermediate case, where the approximation remains close but exhibits a larger gap at higher QP values. Netflix Crosswalk is among the most challenging sequences in the dataset: ApproxSSIMate follows the measured trend at high and medium quality, but increasingly underestimates measured SSIM under stronger compression. This sequence-level view confirms the main limitation observed in the aggregate and boxplot results: the approximation remains trend-consistent, but its absolute error can increase for challenging content at very low quality.

Overall, these results indicate that ApproxSSIMate provides a reliable low-cost estimate of sequence-level SSIM trends across three encoder families. It is most accurate in the high- and medium-quality range, while very low-quality encodes remain the main limitation.

## Discussion

The main practical advantage of ApproxSSIMate is that it changes where the cost of quality estimation is paid. In a conventional workflow, each candidate encode must be decoded and compared against the source sequence to compute SSIM. In the proposed workflow, the source-dependent statistics are computed once and reused across all candidate encodes of the same asset, while the distortion-dependent term is obtained from MSE or PSNR measurements that are already available in many encoding pipelines. This makes the method particularly suitable for per-title encoding, shot-based optimization, and context-adaptive workflows where many encodes are evaluated for the same source content.

The main practical advantage of ApproxSSIMate is that it changes where the cost of quality estimation is paid. In conventional workflows, each candidate encode must be decoded and compared against the source sequence to compute SSIM. In the proposed workflow, source-dependent statistics are computed once and reused, while the distortion-dependent term is obtained from MSE or PSNR measurements already available in many encoding pipelines. This makes the method suitable for per-title encoding, shot-based optimization, and context-adaptive workflows.

ApproxSSIMate should therefore be interpreted as a lightweight estimator rather than a replacement for full-reference quality assessment. Its value is highest when the objective is to compare many candidate encodes, prune the search space, or provide a fast quality signal inside an adaptive encoding loop. Full SSIM, VMAF, or subjective assessment may still be appropriate for final validation, product-level quality gates, or cases where small quality differences have significant operational impact.

The method also has limitations. Because the approximation redistributes global distortion according to reference-side spatial statistics, it is best suited to compression-like distortions where global MSE remains informative. Accuracy may degrade at very low bitrates, for strongly localized artifacts, or for distortions that alter local structure in ways not captured by the reference-derived statistics.

## CONCLUSION

Perceptual metrics such as SSIM are not only useful in modern video pipelines; they are increasingly part of the decision-making process of those pipelines. However, computing perceptual metrics is not free. The cost becomes significant when quality measurements are repeated across many candidate encodes at scale.

ApproxSSIMate is a fast, sequence-level SSIM approximation derived from reference-side spatial statistics and encoder-available distortion information such as MSE or PSNR. Experiments on the Objective-1-fast dataset using x264, x265, and SVT-AV1 show that ApproxSSIMate closely tracks measured mean SSIM over most of the tested quality range.

The main modelling limitation is observed at very low quality, where the approximation error becomes more content-dependent and tends to underestimate measured SSIM. More generally, ApproxSSIMate shows a small positive bias for high-quality encodes and a negative bias for low-quality encodes, while preserving the monotonic degradation trend of SSIM as QP increases.

In addition, the current implementation is a Python research prototype. For widespread deployment, the next step is an optimized C or C++ implementation, including SIMD acceleration for the reference-side window statistics and integration with existing encoder or quality-metric pipelines. ApproxSSIMate should therefore be viewed as a fast decision-support signal for low-latency streaming, operating-point comparison, and content-adaptive optimization, rather than as a universal replacement for full-reference quality assessment.

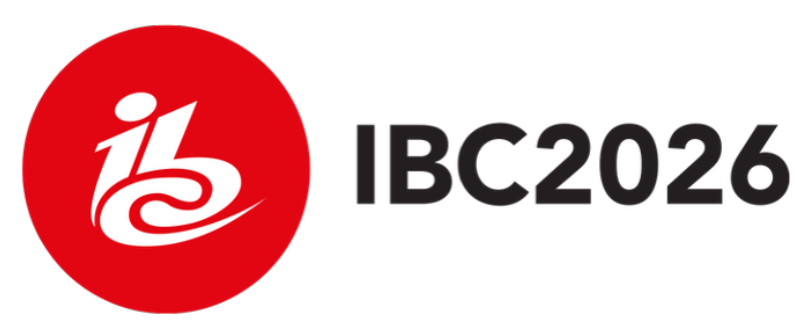